\documentclass{iopart}
\usepackage{graphics,graphicx}
\begin{document}
\title[Medium Modifications of Charm and Charmonium in HI
collisions]{Medium Modifications of Charm and Charmonium in
  High-Energy Heavy-Ion Collisions} 
\author{L Grandchamp$^1$, R Rapp$^2$ and G E Brown$^3$}
\address{$^1$\ Lawrence Berkeley National Laboratory, Berkeley, CA 94720}
\address{$^2$\ Cyclotron Institute and Department of Physics, 
Texas A\&M University, College Station, TX 77843}
\address{$^3$\ Department of Physics, SUNY Stony Brook, NY 11794}
\ead{LGrandchamp@lbl.gov, rapp@comp.tamu.edu}
\begin{abstract}
The production of charmonia in heavy-ion collisions is 
investigated within a kinetic theory framework simultaneously 
accounting for dissociation and regeneration processes  
in both quark-gluon plasma (QGP) and hadron-gas phases of
the reaction. In-medium modifications of open-charm states 
($c$-quarks, $D$-mesons) and the survival of $J/\psi$ mesons 
in the QGP are included as inferred from lattice QCD. 
Pertinent consequences on equilibrium charmonium abundances are 
evaluated and found to be especially relevant to explain the 
measured centrality dependence of the $\psi'/\psi$ ratio at SPS. 
Predictions for recent $In$-$In$ experiments, as well as 
comparisons to current $Au$-$Au$ data from RHIC, are provided.  
\end{abstract}

\section{Introduction}          
Heavy-flavor bound states constitute a valuable probe of the hot/dense
strongly interacting matter formed in relativistic collisions of 
heavy nuclei. It was first suggested in \cite{Matsui:1986dk} to
identify $J/\psi$ suppression as a signature of a deconfined medium 
due to color Debye screening, 
with tightly bound $c\bar{c}$ states presumably being robust  
in a hadron gas (HG). However, more recently it has been 
realized that $c$-quark reinteractions in the medium can lead to 
regeneration of charmonium states through $c$-$\bar c$ 
coalescence~\cite{Braun-Munzinger:2000px,Thews:2000rj}, 
especially if charm production is abundant
(\textit{e.g.}, $N_{c\bar{c}}$$\sim$10-20 in central $Au$-$Au$
collisions at RHIC).

Further insights into charm(onium) properties at finite temperature 
$T$ have re\-cently 
been provided by lattice QCD (LQCD) calculations which indicate
\textit{(i)} a continuous reduction of the open-charm threshold with  
increasing matter tempera\-ture~\cite{Karsch:2000kv} and
\textit{(ii)} the survival of low-lying charmonia ($\eta_c, J/\psi$)
up to $\sim 2\;T_c$ \cite{Datta:2002ck,Umeda:2002vr,Asakawa:2003re}.

We here present an approach to charmonium production at SPS and 
RHIC~\cite{Grandchamp:2003uw} in which in-medium charm 
properties are modeled in accord with LQCD results and implemented into
a kinetic rate equation, solved for a schematic thermal fireball
expansion~\cite{Rapp:1999ej}. It enables a simultaneous treatment of 
charmonium dissociation and regeneration throughout the evolution of the 
system, thus improving on our earlier constructed ``two-component'' 
model~\cite{Grandchamp:2001pf,Grandchamp:2002wp} where
suppression in QGP and HG was combined with statistical production at
hadronization.

\section{In-medium properties of open and hidden charm in equilibrium}
LQCD calculations at finite $T$ show that the free energy
of a static heavy-quark pair ($Q\bar Q$), $F_{Q\bar Q}$=$V$-$TS$, 
reaches a (constant) plateau at large spatial separation~\cite{Karsch:2000kv},
which decreases with increasing $T$, continuously across the phase 
transition. We associate this behavior with a decreasing open-charm 
threshold\footnote[1]{Note that for a quantitative assessment,  
the entropy term in $F_{Q\bar Q}$ should be removed.}; in the  
hadronic phase, a plausible interpretation is provided by the   
(partial) restoration of chiral symmetry: as $T$
increases, we expect a reduction of constituent light-quark masses,
inducing smaller $D$-meson masses. The evaluation of the former,   
within a NJL model at finite temperature and density, typically amounts 
to $\Delta m(T_c) \simeq -140$ MeV. Requiring continuity across $T_c$,
the effective charm-quark mass in the QGP is around 
$m_c^*\simeq$~1.6-1.7~GeV, which might be associated with 
a thermal correlation energy of $c$-quarks. 

At a fixed number, $N_{c\bar c}$, of $c\bar c$ pairs in the system
(as expected for production exclusively in initial $N$-$N$ collisions), 
in-medium modified open-charm states affect the (thermal) 
equilibrium abundance of charmonia $\Psi$, given by the densities 
$n_{\Psi}^{\rm{eq}}(T,\gamma_c) = d_\Psi \gamma_c^2 \int
\frac{d^3q}{(2\pi)^3}f^{\Psi}(m_{\psi},T)$, 
where $f^{\Psi}$ are Bose distribution functions (using vacuum $\Psi$ 
masses), and $d_\Psi$ is a spin-isospin degeneracy. The charm-quark 
fugacity $\gamma_c$ encodes chemical 
off-equilibrium effects being adjusted to $N_{c\bar{c}}$ through
\begin{equation}
  \label{eq:2}
N_{c\bar{c}} = \frac{1}{2}\gamma_c
N_{\rm{op}}\frac{I_1(\gamma_cN_{\rm{op}})}{I_0(\gamma_cN_{\rm{op}})} +
V\sum\limits_{\Psi=\eta_c,J/\psi,\cdots}^{} n_{\Psi}^{\rm{eq}}(T,\gamma_c)\ ,  
\end{equation}
where $N_{\rm{op}}=Vn_{\rm{op}}(m_{c,D}^*;T)$ denotes the total equilibrium
number of open-charm states in either QGP ($c$, $\bar{c}$
quarks) or HG phase (charmed hadrons). In the QGP, 
an in-medium $c$-quark mass, larger than the perturbative value
of $m_c \sim 1.2$ GeV, leads to an increase in $\Psi$ equilibrium
abundances since it becomes energetically favorable to distribute 
$c\bar c$ pairs into $\Psi$ states, especially if $m_{\Psi} <
2m_c^*$. Conversely, in the hadronic phase, reduced in-medium
$D$-meson masses lower the $\Psi$ equilibrium level. 

\section{Rate equations in heavy-ion collisions}
To make contact with the dynamical situation encountered in heavy-ion
collisions, we employ a kinetic theory framework. If open-charm states 
are in thermal equilibrium, the evolution equation for the number, 
$N_{\Psi}$, of charmonium states present in the system is given by
$\frac{dN_{\Psi}}{d\tau}=-\frac{1}{\tau_{\Psi}}[N_{\Psi}-N_{\Psi}^{\rm{eq}}]$.
The charmonium equilibrium numbers $N_{\Psi}^{\rm{eq}}$ are determined 
as described above, except, however, for additional thermal 
off-equilibrium corrections expected in heavy-ion collisions, namely:  
\textit{(i)} a thermal relaxation time which effectively reduces 
$N_{\Psi}^{\rm{eq}}$ during the early times of the collision (see 
\cite{Grandchamp:2002wp} for details), and \textit{(ii)} a correlation volume 
$V_{\rm{corr}}$ implemented into the argument of the Bessel functions 
in Eq.~(\ref{eq:2}) to account for the locality of 
$c$-$\bar c$ production. The thermal widths of charmonia, 
$\Gamma_{\Psi}=(\tau_{\Psi})^{-1}$, are obtained by convoluting 
their inelastic cross sections with thermal distributions of the 
matter constituents. In the QGP phase, we use parton-induced 
``quasifree'' breakup reactions~\cite{Grandchamp:2001pf},
 $i+\Psi\rightarrow i+c+\bar{c}$ ($i=g,q,\bar{q}$), accounting for 
the different binding energies of the respective charmonium states 
($\psi,\psi'$ and $\chi$).  In the HG, we compute $J/\psi$ 
breakup by pions and rhos within a flavor-$SU(4)$ effective lagrangian
formalism (see \cite{Duraes:2002ux} and references therein),
augmented by geometric scaling for the $\psi'$ and $\chi$. 
Here, the consequences of in-medium $D$-meson masses are two-fold: 
first, inelastic reaction rates increase due to a larger
available phase space and second, the $D\bar{D}$ threshold can move
below the charmonium mass, which additionally enables direct decays
$\Psi\rightarrow D\bar{D}$~\cite{Friman:2002fs}. 
The rate equations are supplemented by initial conditions,
$N_{\Psi}^0$, for which we take experimental yields from
$pp$ extrapolated in centrality by $N$-$N$ collision scaling and 
``pre-equilibrium'' nuclear absorption. 
The rate equations are then 
integrated over the space-time history of the collision according to 
a schematic thermal fireball evolution~\cite{Rapp:1999ej}. 

\section{Comparison with experiments}
\begin{figure}[!t]
\centering
\includegraphics[width=0.45\textwidth,clip=]{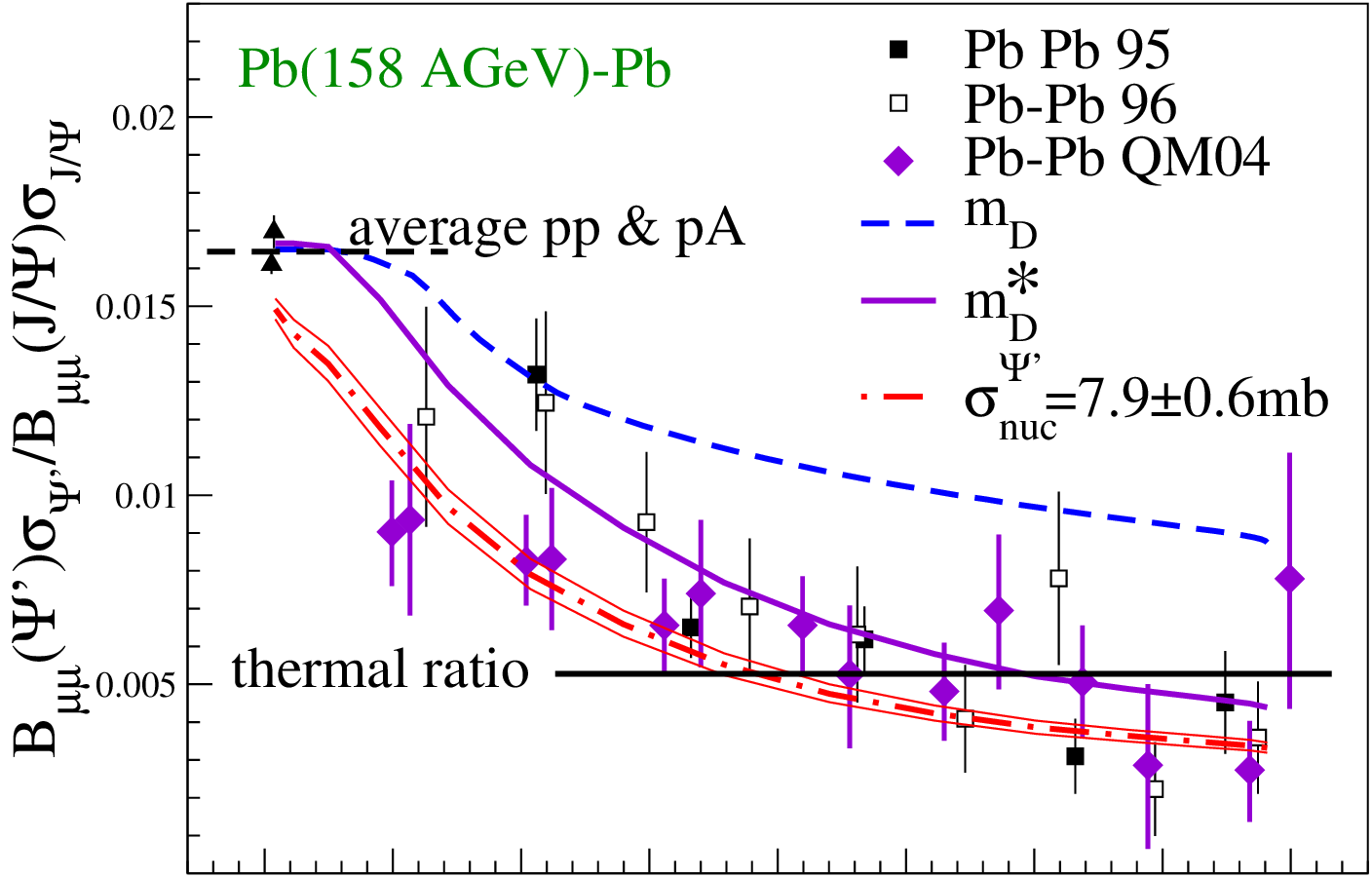}
\hspace{0.5cm}
\includegraphics[width=0.45\textwidth,clip=]{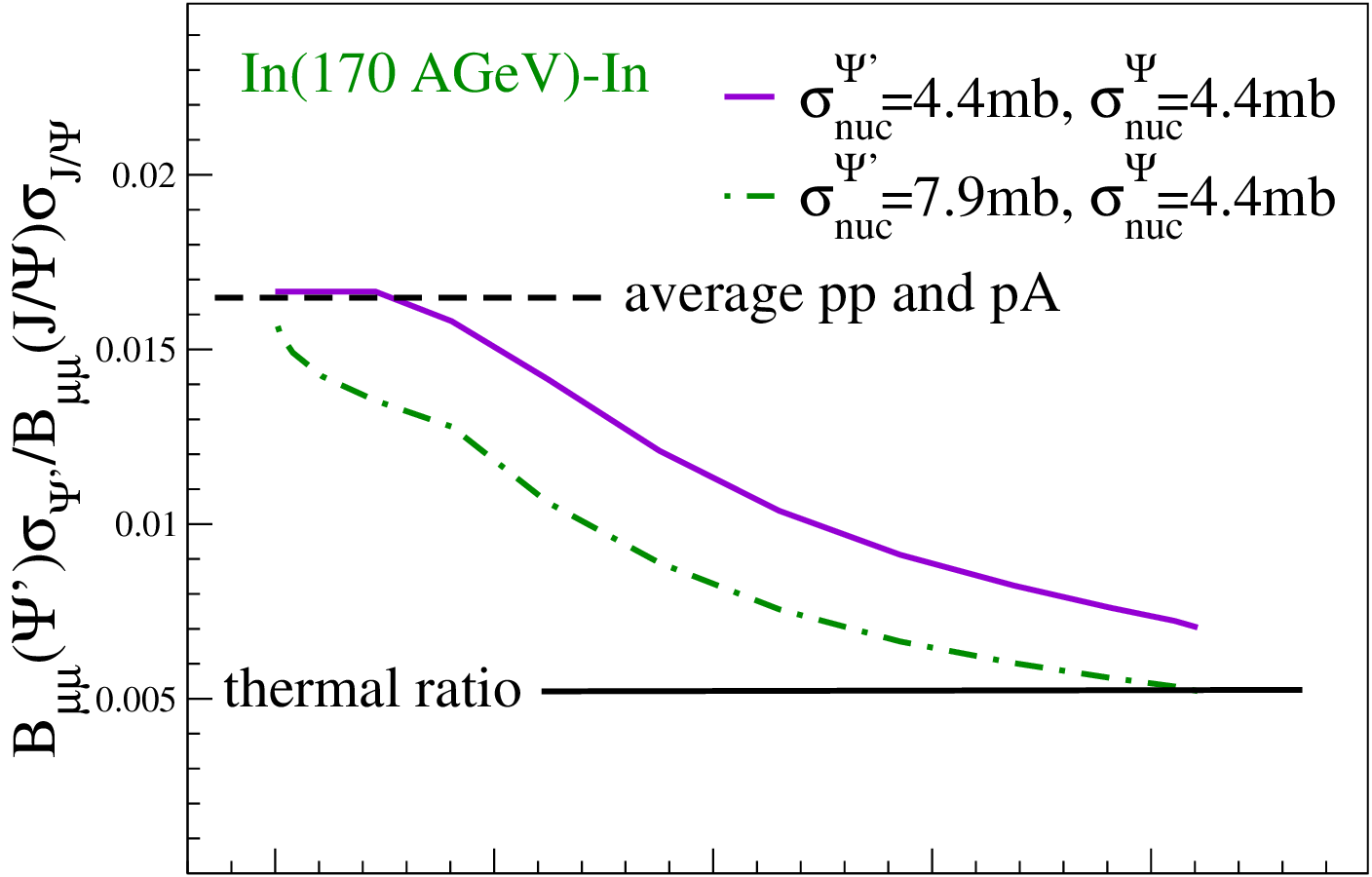}
\\
\vspace*{-0.06cm}
\includegraphics[width=0.45\textwidth,clip=]{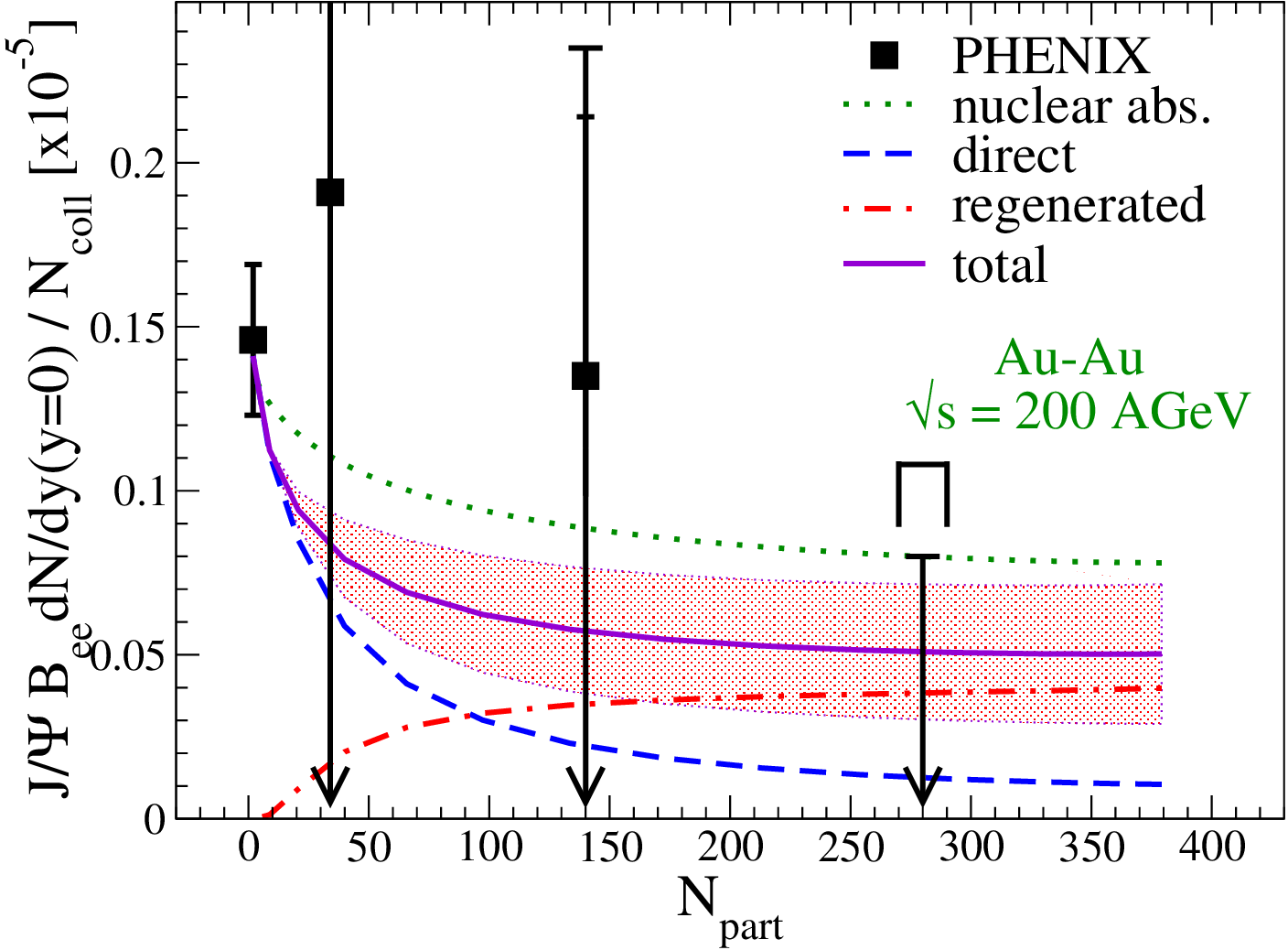}
\hspace{0.5cm}
\includegraphics[width=0.45\textwidth,clip=]{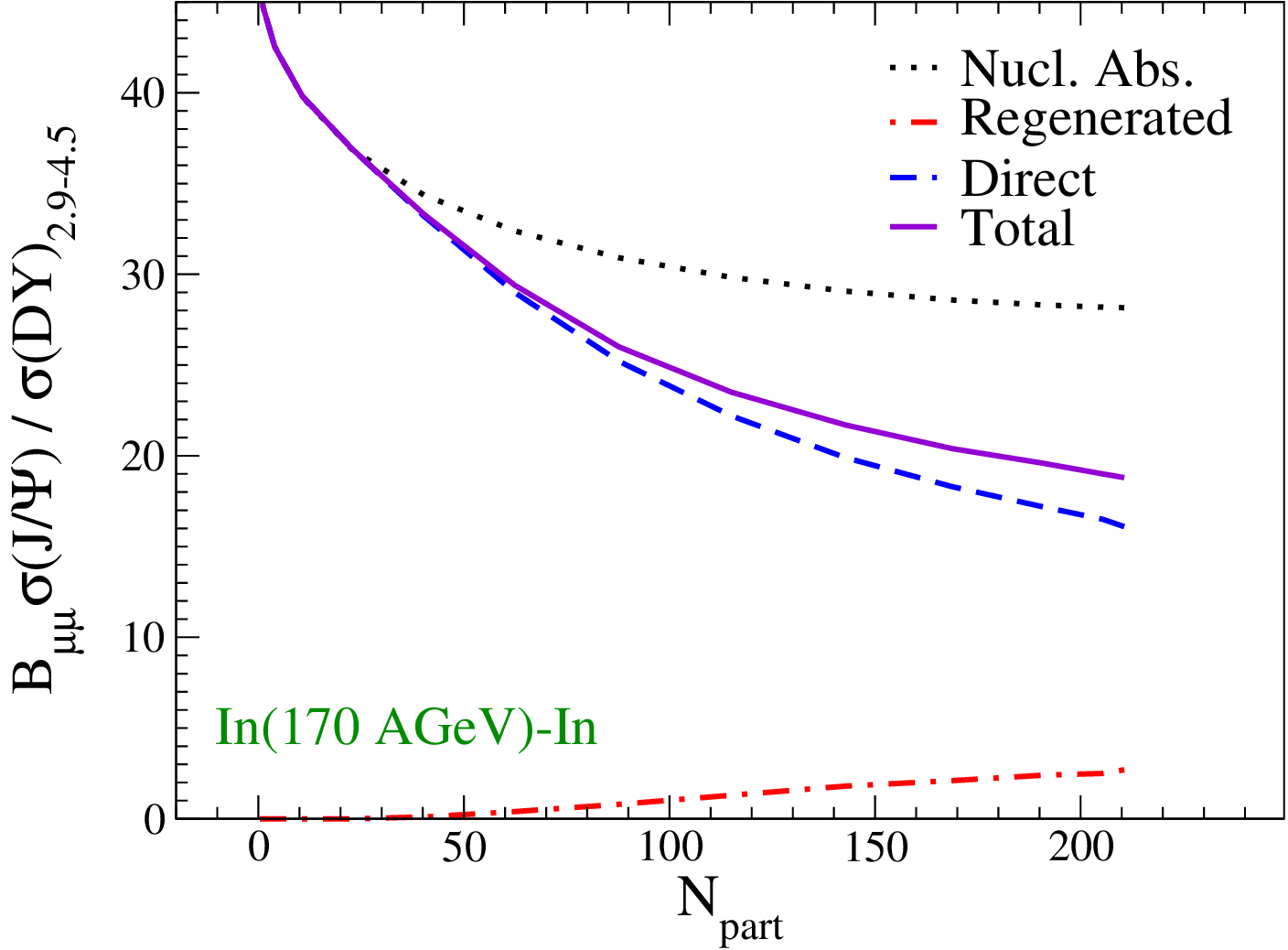}
\caption{Upper left panel: $\psi'/\psi$ ratio in $Pb$(158AGeV)-$Pb$ at SPS; 
solid (dashed) line:  with (without) in-medium effects using a uniform nuclear
absorption cross section $\sigma_{\rm{nuc}}=4.4$ mb; dash-dotted line:
as solid line but with an updated value for the $\psi'$ nuclear 
absorption cross section, $\sigma_{\rm{nuc}}^{\psi'}=7.9\pm0.6$~mb. 
Lower left panel: $J/\psi/N_{\rm{coll}}$ at mid-rapidity vs. $N_{\rm{part}}$
   in 200~AGeV $Au$-$Au$ at RHIC. The total yield (full
  curve) is dominated by regenerated $J/\psi$'s (dash-dotted curve)
  with almost no direct $J/\psi$'s remaining (dashed curve). The band
  reflects the uncertainty on the magnitude of in-medium
  effects. Upper (lower) right panel: centrality dependence of the 
  $\psi'/\psi$ ($J/\psi/DY$) ratio in $In$(170 AGeV)-$In$ at SPS.}
\label{fig:sps-PbPb}
\end{figure}
\textit{Pb-Pb at SPS} $-$ At SPS energies ($\sqrt{s_{NN}}$=17.3~GeV), 
primordial charmonium production is large compared to charmonium 
equilibrium abundances, $N_{\psi}^0\gg N_{\psi}^{eq}$, implying little 
regeneration. As shown in \cite{Grandchamp:2003uw}, the centrality
dependence of the $J/\psi$ over Drell-Yan ratio in $Pb$-$Pb$
collisions is well reproduced and QGP formation is characterized by 
$J/\psi$ suppression. The consequences of in-medium effects 
at SPS are particularly pronounced 
in the $\psi'/\psi$ ratio, cf.~the upper left panel of 
Fig.~\ref{fig:sps-PbPb}. With vacuum $D$-meson masses (dashed line)
our calculation underestimates $\psi'$ suppression. The calculation 
including medium effects (full line) improves the agreement with NA50 
data~\cite{Abreu:1998vw} substantially, which is a direct consequence 
of the reduction of the $D\bar{D}$ threshold in the HG, opening the 
$\psi'\rightarrow D\bar{D}$ decay channel.  The $\psi'$ data set
(including $p$-$A$ collisions) has 
recently been reanalyzed by NA50~\cite{Santos:2004} 
(diamonds), deducing a stronger nuclear absorption of the 
$\psi'$, $\sigma_{\rm{nuc}}(\psi')=7.9 \pm 0.6$~mb $>$
$\sigma_{\rm{nuc}}(\psi)=4.4$ mb. Our calculation with the 
correspondingly updated values of $\sigma_{\rm{nuc}}$ is shown 
by the dash-dotted line, confirming the need for in-medium effects 
to reproduce the $\psi'/\psi$ ratio. 

\textit{Predictions for In-In at SPS} $-$ The NA60 collaboration will
continue to study charm(onium)  production at SPS and we 
present in the right panel of Fig.~\ref{fig:sps-PbPb} our predictions
for the $\psi'/\psi$ and $J/\psi/DY$ ratios as
a function of centrality in $In$(170 AGeV)-$In$ collisions. 
Similar to the $Pb$-$Pb$ system, $J/\psi$ regeneration (dash-dotted
curve in the lower right plot) is rather limited and the main effect remains
$J/\psi$ suppression (dashed curve) exhibiting a marked departure
from nuclear absorption (dotted line). Medium effects are
still appreciable, as illustrated in the upper right panel where a
gradual decrease of the $\psi'/\psi$ ratio is predicted as a function of
$N_{\rm{part}}$.

\textit{RHIC results} $-$ Our calculations at full RHIC energy 
($\sqrt{s_{NN}}=200$~GeV) are compared to published PHENIX
data~\cite{Adler:2003rc} in the lower left of Fig.~\ref{fig:sps-PbPb}.
Contrary to SPS, the $J/\psi$ yield in
central $Au$-$Au$ collisions (full curve) is dominated by regenerated $J/\psi$'s
(dash-dotted curve) while primordial $J/\psi$'s are almost completely
suppressed (dashed line). The uncertainty linked to our treatment of 
in-medium effects is reflected by the band corresponding to 
$-250 < \Delta m_D(T_c) < -80$ MeV, with stronger in-medium
effects resulting in a smaller $J/\psi$ yield. 

\section{Conclusions}
We have presented a model for charmonium production in heavy-ion
collisions incorporating in-medium effects on open-charm states, as 
inferred from Lattice QCD, within a kinetic rate equation which allows 
to comprehensively treat suppression and regeneration mechanisms during 
the course of the collision. We have found that QGP formation manifests 
itself by $J/\psi$ suppression at SPS energies and by $J/\psi$ 
regeneration at RHIC, where run-4 data are expected to give important 
insights. In-medium effects have so far proved to be essential to 
understand the centrality dependence of the $\psi'/\psi$ 
ratio at SPS. Our predictions can also be tested by upcoming NA60 
data for $In$-$In$ collisions at SPS. Complementary studies of
charmonium transverse momentum distributions, as well as charmonium and 
bottomonium production at LHC, will provide further scrutiny
of the proposed approach.

\section*{References}
\bibliographystyle{JPhysG}
\bibliography{qm04-proc}

\begin{thebibliography}{10}
\providecommand{\url}[1]{\texttt{#1}}
\providecommand{\urlprefix}{URL }
\providecommand{\bibAnnoteFile}[1]{%
  \IfFileExists{#1}{\begin{quotation}\noindent\textsc{Key:} #1\\
  \textsc{Annotation:}\ \input{#1}\end{quotation}}{}}
\providecommand{\bibAnnote}[2]{%
  \begin{quotation}\noindent\textsc{Key:} #1\\
  \textsc{Annotation:}\ #2\end{quotation}}
\providecommand{\eprint}[2][]{{\em{Preprint}} #2}

\bibitem{Matsui:1986dk}
Matsui T and Satz H 1986 \emph{Phys. Lett.} B \textbf{178} 416
\bibAnnoteFile{Matsui:1986dk}

\bibitem{Braun-Munzinger:2000px}
Braun-Munzinger P and Stachel J 2000 \emph{Phys. Lett.} B \textbf{490} 196
\newblock \eprint{nucl-th/0007059}
\bibAnnoteFile{Braun-Munzinger:2000px}

\bibitem{Thews:2000rj}
Thews R~L, Schroedter M and Rafelski J 2001 \emph{Phys. Rev.} C \textbf{63}
  054905
\newblock \eprint{hep-ph/0007323}
\bibAnnoteFile{Thews:2000rj}

\bibitem{Karsch:2000kv}
Karsch F, Laermann E and Peikert A 2001 \emph{Nucl. Phys.} B \textbf{605} 579
\newblock \eprint{hep-lat/0012023}
\bibAnnoteFile{Karsch:2000kv}

\bibitem{Datta:2002ck}
Datta S, Karsch F, Petreczky P and Wetzorke I 2002  \eprint{hep-lat/0208012}
\bibAnnoteFile{Datta:2002ck}

\bibitem{Umeda:2002vr}
Umeda T, Nomura K and Matsufuru H 2002  \eprint{hep-lat/0211003}
\bibAnnoteFile{Umeda:2002vr}

\bibitem{Asakawa:2003re}
Asakawa M and Hatsuda T 2004 \emph{Phys. Rev. Lett.} \textbf{92} 012001
\newblock \eprint{hep-lat/0308034}
\bibAnnoteFile{Asakawa:2003re}

\bibitem{Grandchamp:2003uw}
Grandchamp L, Rapp R and Brown G~E 2003  \eprint{hep-ph/0306077}
\bibAnnoteFile{Grandchamp:2003uw}

\bibitem{Rapp:1999ej}
Rapp R and Wambach J 2000 \emph{Adv. Nucl. Phys.} \textbf{25} 1
\newblock \eprint{hep-ph/9909229}
\bibAnnoteFile{Rapp:1999ej}

\bibitem{Grandchamp:2001pf}
Grandchamp L and Rapp R 2001 \emph{Phys. Lett.} B \textbf{523} 60
\newblock \eprint{hep-ph/0103124}
\bibAnnoteFile{Grandchamp:2001pf}

\bibitem{Grandchamp:2002wp}
Grandchamp L and Rapp R 2002 \emph{Nucl. Phys.} A \textbf{709} 415
\newblock \eprint{hep-ph/0205305}
\bibAnnoteFile{Grandchamp:2002wp}

\bibitem{Duraes:2002ux}
Duraes F~O, Kim H~C, Lee S~H, Navarra F~S and Nielsen M 2002
  \eprint{nucl-th/0211092}
\bibAnnoteFile{Duraes:2002ux}

\bibitem{Friman:2002fs}
Friman B, Lee S~H and Song T 2002 \emph{Phys. Lett.} B \textbf{548} 153
\newblock \eprint{nucl-th/0207006}
\bibAnnoteFile{Friman:2002fs}

\bibitem{Abreu:1998vw}
Abreu M~C \emph{et~al} (NA50 Collaboration) 1998 \emph{Nucl. Phys.} A
  \textbf{638} 261
\bibAnnoteFile{Abreu:1998vw}

\bibitem{Santos:2004}
Santos H (NA50 Collaboration) 2004  these proceedings
\bibAnnoteFile{Santos:2004}

\bibitem{Adler:2003rc}
Adler S~S \emph{et~al} (PHENIX Collaboration) 2003  \eprint{nucl-ex/0305030}
\bibAnnoteFile{Adler:2003rc}

\end{thebibliography}

\end{document}